# Anisotropic polarization, predicted as a result of the diffraction of blackbody radiation at a reflective phase grating with ideal conductivity

V. V. Savukov

In the course of analyzing the axiomatic principles that form the basis of statistical physics, the validity of the postulate that all the isoenergetic microstates of a closed system are equally probable was checked. This article reports the results of numerically modelling the interaction of thermodynamically equilibrium blackbody radiation with a reflective phase diffraction grating that possesses ideal conductivity. Cases are found in which anisotropy of the polarization parameters is guaranteed to appear inside a closed volume of initially homogeneous blackbody radiation, resulting in a formal decrease of its Boltzmann entropy as a consequence of deviation from the microcanonical Gibbs distribution. This is apparently caused by the discontinuous character of the change of the phase trajectories of the photons during diffraction, which makes the physical system under consideration nonergodic.

*St. Petersburg, Russia   2012*

# TABLE OF CONTENTS

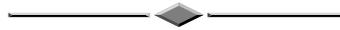

# Introduction

The existing apparatus of the statistical physics of equilibrium systems is based on the hypothesis that all microstates accessible to the closed system under consideration are equally probable. This implies the following:

1. The most probable stationary state of a closed physical system (isolated from the environment) is called the equilibrium state. The equilibrium state is macroscopic. It is the collection of all the microstates accessible to the system—i.e., of such specific states each of which can be realized at a given energy level.

2. The equilibrium state is realized at each fixed instant via one of the components of its microstates. The system in this case can be detected with identical probability in any of the microstates that form its equilibrium macrostate.

Point 2 of the given axiomatics makes it possible to declare just how the equilibrium state of a closed system will look and also gives a basis for determining the direction of any stochastic processes in time. This means, for instance, that the diffraction scattering of a photon gas observed in such a system must not alter the macroscopic parameters of this gas at all if they already correspond to the definition of the equilibrium state described above.

Meanwhile, a diffraction-scattering mechanism of a gas of quantum particles makes it theoretically allowable for nonergodic systems[1] to exist whose behavior lies outside the "zone of accountability" of statistical physics. This is because the phase trajectory of each quantum particle is discontinuous at the level of momentum subspace[2]. Moreover, such a trajectory cannot be described even by a discontinuous functional dependence that obeys the Dirichlet conditions: If each diffraction-scattering event of a quantum particle is interpreted as a discontinuity of the first type in the momentum subspace of its phase $\mu$-space, an unambiguously defined[3] pure state of the particle before scattering, for example, will correspond to some probability factor of the states after scattering. Moreover, according to the Copenhagen school, there are no hidden parameters that make it possible to eliminate the indicated ambiguity of the trajectory function.

The described capability of quantum particles to appear and disappear at various parts of phase space accessible to them opens up the possibility that sources and sinks of the phase trajectories that have a nonidentical density in the same local sections of the phase volume exist in this space. The time-stable nonzero divergence of the flow of phase trajectories that appears as a result in specific parts of phase space can make the given system nonergodic, while its properties are made incompatible with the axiomatics of statistical physics [1].

---

[1] The property of ergodicity assumes the validity of the microcanonical hypothesis of statistical physics that the results of averaging over time and phase are identical when the values of the macroscopic parameters of a system are computed.

[2] As applied to quantum particles, the concept of phase trajectory can be preserved by redefining it on the basis of Ehrenfest's theorem, which describes the dynamics of the center of gravity of a volume of "probability fluid" with density $\rho = |\Psi|^2$, normalized to unity. In this case, one usually speaks not so much of a phase trajectory as of a trajectory tube, whose width depends on the probability "fuzziness" of the corresponding coupled parameters of the particle.

[3] Of course, only to within Heisenberg's uncertainty.

What has been said gives a basis for assuming that, for example, a diffuse monochromatic photon gas[1] that experiences "elastic scattering" (with no losses) at an ideally conductive reflecting phase diffraction grating can change its initial isotropic macrostates to an anisotropic state. In other words, the microstates of the photons of such a gas, initially uniformly distributed inside some isoenergetic layer of phase space, can be nonuniformly redistributed in the volume of the given layer as a consequence of the diffraction process at an ideally reflecting grating. If a diffuse photonic gas and a diffraction grating are component parts of a closed physical system, the Boltzmann entropy of this system in the course of such a process must vary arbitrarily.

This article presents the results of a verification of this assumption, carried out by means of numerical modelling.

## Formulation of the problem

Let a limited volume of some closed physical system be filled with the thermodynamically equilibrium radiation of an absolute blackbody having a Planckian frequency spectrum. An ideally conductive (photon-nonabsorbing) reflective diffraction grating interacting with the given radiation is placed in the indicated volume.

It follows from the axiomatic postulates of statistical physics mentioned earlier that none of the isoenergetic scattering processes must have any effect on the values of the macroscopic parameters of the blackbody radiation as a whole or of its separate monochromatic components in frequency, since these values are the most probable for the current energy level of the system under consideration. A different result would mean that there were limits to the correct application of the indicated postulates with respect to quantum systems in the uncondensed state.

## Methodology of predicting the states

An isolated physical system that contains a stochasticized diffuse photon gas that interacts with a special reflective optical element was chosen as the object of simulation modelling. This element, in particular, can be a phase-type reflective diffraction grating with one-dimensional (linear) sinusoidal surface microrelief.

The macroscopic parameters of the total light field, formed as a result of the given interaction, are computed on the basis of the corresponding theoretical model. In carrying out this project, software was specially created and tested for simulation modelling of the scattering processes of a diffuse photon gas on various forms of reflective optical elements (see Refs. [2-5] for details).

The diffuse structure of the initial light field is generated by means of selections of the initial parameters valid in probability for each wave object associated with a single photon. The results of the solutions obtained for elementary scattering events of the individual photons at the optical element are stored by the program. The cumulative parameters of the radiation are found in numerical form as the statistical moments of the distributions obtained by generalizing the information of the calculations for all the single elementary scattering events.

---

[1] By a diffuse photon gas is meant here unpolarized incoherent electromagnetic radiation, for the individual photons of which any possible angular orientation of their wave vectors **k** is equally probable in three-dimensional geometrical space.

To make it more obvious, the cumulative information is also displayed as graphic images—for example, in the form of angular diagrams of the radiance of the scattered radiation.

## Reflection from ideally conducting surfaces

The problem of radiation scattering at ideally conducting reflecting surfaces is characterized by the following distinguishing features:

● Because there is no absorption and no intrinsic thermal emission of photons by such a surface, the scattering process has an isoenergetic character. This circumstance makes it possible to introduce such a process into a closed physical system without changing its energy level. It is interesting that the internal volume of the ideally reflective element does not formally enter into the composition of the given system and can, for example, have an arbitrary temperature.

● The fact that it is unnecessary to consider processes of intrinsic thermal radiation of microstructured surfaces greatly simplifies the formulated problem, although it should be assumed that, in the presence of such processes, it is possible to obtain extremely nontrivial results, associated with the assumed deviations from Planck's law [6] and Lambert's law [7]. We recall that Lambert's law, applied to the total radiance of the radiation emitted and scattered by the interior surface of a closed system, is one version of the boundary condition that determines the validity of the second law of thermodynamics [8].

● When a photon gas is scattered at an ideally conductive reflecting surface, Planck's and Kirchhoff's laws will be obeyed automatically, since their validity in this case directly follows from the law of conservation of energy. The second principle of thermodynamics, in turn, limits the possible anisotropy of the scattered gas to macroparameters that do not depend on the radiance.

Figure 1 shows images of the angular distributions of the radiance of various polarization components of the scattered light field (S+P, S, P), produced while verifying the program of Refs. [2, 3] and obtained for a mirror-smooth reflective surface with ideal conductivity. The initial light field is monochromatic (wavelength $\lambda = 555$ nm) diffuse radiation with a total number of photons in the statistical experiment of $2^{20}$ - 1 = 1048575. Each angular distribution is constructed in a polar coordinate system in such a way that its center corresponds to a zero value of the angle of reflection for an external view of the surface. The angle of reflection is proportional to the polar radius, and this angle approaches 90° at the periphery of the circular diagram. The azimuthal angle of observation of the mirror surface is determined by the polar angle of the diagram.

The separate diagrams are automatically scaled so that all the existing density contrasts of the scattered light flux are maximized. It is obvious that only random manifestations of the fluctuations of the scattered field, which do not form any visually observable macroscopic gradients, are present on the images of the angular distributions in Figs. 1(a)– 1(c). In other words, when diffuse radiation is reflected from an ideally conducting mirror, Lambert's law is obeyed for all scattered components, as is quite expected [9].

Each of the circular diagrams is accompanied by a graph (Figs. 1(d)– 1(f)) of their horizontal semicross-sections (zero azimuthal angle), giving a representation of the actual radiance scale in these planes with respect to the Lambertian isotropic level, denoted by a dotted line for 1 (Fig. 1(d)—the sum of the S and P components) or for 1/2 (Figs. 1(e) and 1(f)—the separate S and P components, respectively).

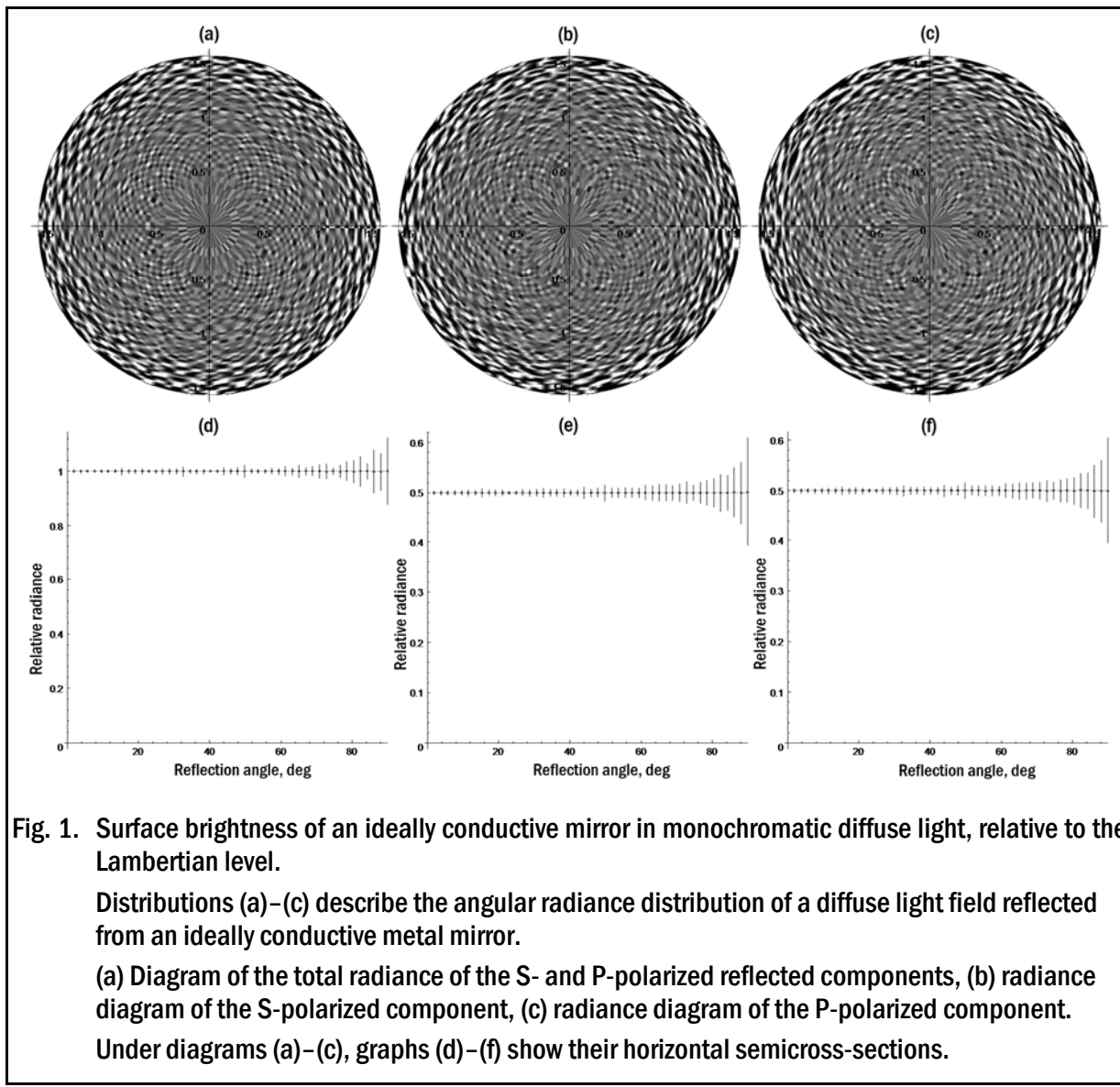

Fig. 1. Surface brightness of an ideally conductive mirror in monochromatic diffuse light, relative to the Lambertian level.

Distributions (a)–(c) describe the angular radiance distribution of a diffuse light field reflected from an ideally conductive metal mirror.

(a) Diagram of the total radiance of the S- and P-polarized reflected components, (b) radiance diagram of the S-polarized component, (c) radiance diagram of the P-polarized component.

Under diagrams (a)–(c), graphs (d)–(f) show their horizontal semicross-sections.

The picture changes if, instead of a mirror, a diffraction grating with the same parameters of the initial light field is taken as the reflective element. Figure 2 shows images of the angular radiance distributions of the polarized components of the scattered light field, obtained for an ideally conductive linear phase grating (step d = 544 nm, total depth of the sinusoidal microrelief profile h = 337 nm, with the microrelief rulings oriented vertically on the diagrams).

The computer-separated macroscopic radiance gradients of the polarized S and P components of the scattered radiation are clearly seen on the diagrams (Figs. 2(b) and 2(c)). The expected radiance variation of the actual physical light field is predicted in this case to be at about 5%–6% of the value corresponding to Lambertian reflection. The radiances of the S and P components always complement each other, so that their sum is ordinary unpolarized diffuse radiation (Fig. 2(a)).

In practice, this effect means the following: Let there be a quasi-closed system in the form of a model of an absolute blackbody (a thermostatically controlled cavity with a small aperture). The existing axiomatics of statistical physics state that, if any material objects placed inside the cavity are in thermodynamic equilibrium with it, an external observer can by no means establish even the fact that these objects are present: Only unpolarized isotropic radiation with a Planck spectrum and a Lambertian angular distribution must always be emitted through the aperture in this case. The result of the simulation experiment, however, is to predict that diffraction objects with a surface that ensures partial or total reflection of the ambient blackbody radiation can be detected inside the cavity. To implement the indicated possibility, a monochromatic component to be analyzed should be separated out from the radiation flux that leaves the cavity, which must then be decomposed into polarization components (see Figs. 2(a)– 2(c)).

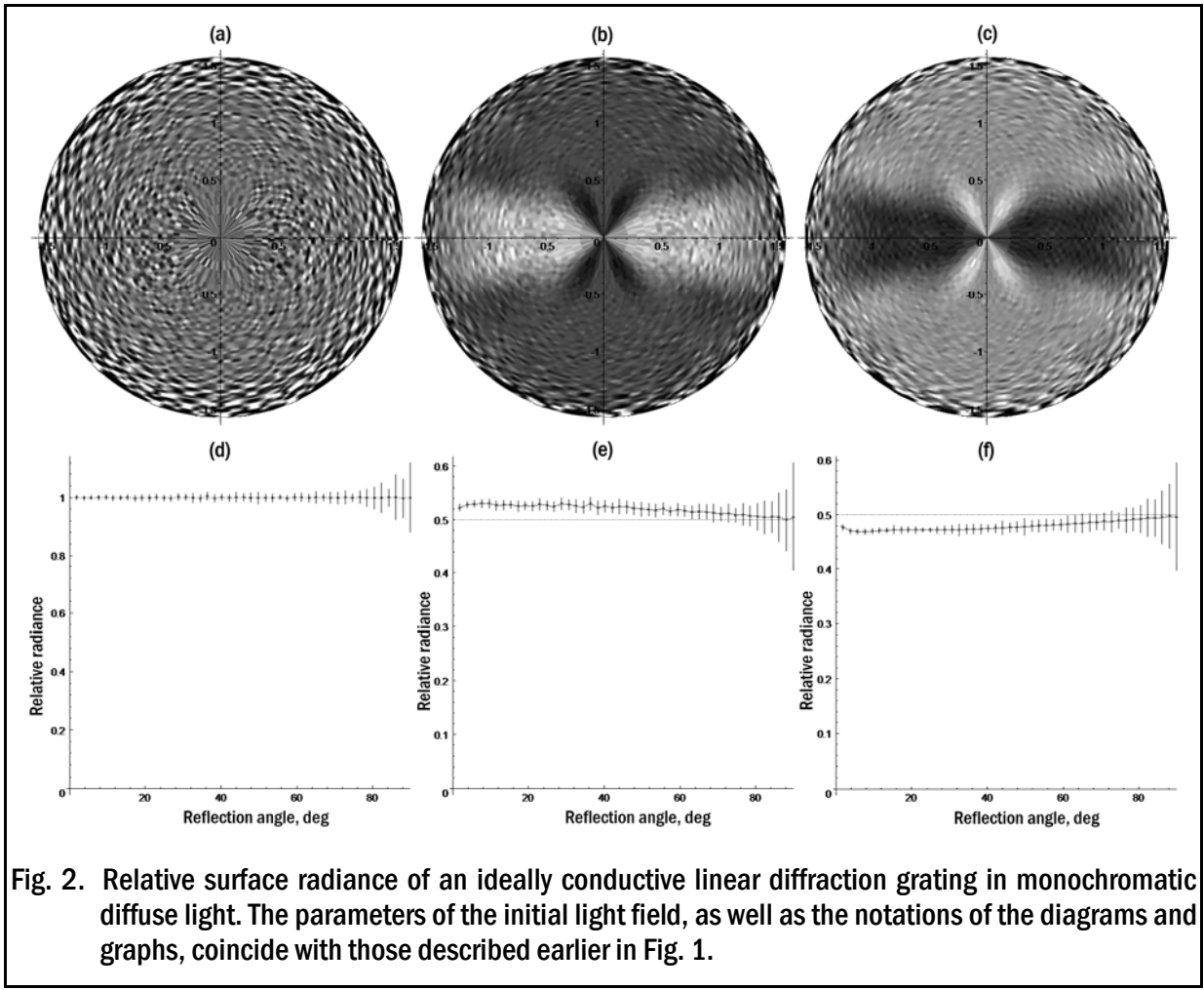

**Fig. 2.** Relative surface radiance of an ideally conductive linear diffraction grating in monochromatic diffuse light. The parameters of the initial light field, as well as the notations of the diagrams and graphs, coincide with those described earlier in Fig. 1.

To check the reliability of this result further, a special verification was carried out in order to eliminate the probability that the detected effect can be explained by the interference of the values of the initial parameters of those photons that form the given diffuse light field. Such interference can theoretically arise on some forms of regular gratings used in the Monte Carlo method when generating the initial characteristics of the individual particles.

Figure 3 shows the same angular radiance distributions as in Fig. 2. The only difference is that a low-dispersion method of Sobol' sequences [10] was used in Fig. 2 to form the network of the initial data of the statistical tests, while the information of Fig. 3 is obtained using for these purposes the standard pseudorandom-number generator included in the **Maple™** mathematical system [2, 3].

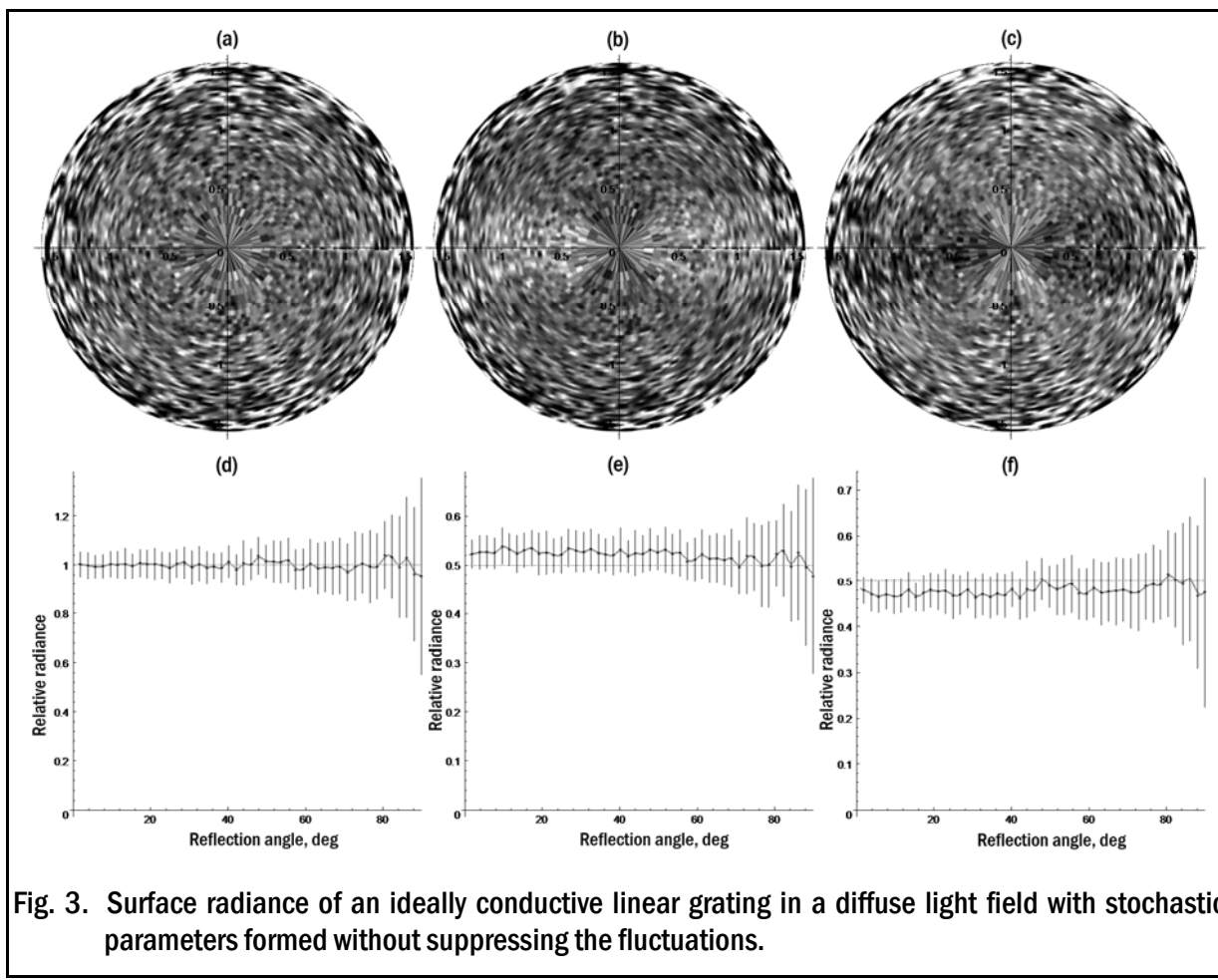

**Fig. 3.** Surface radiance of an ideally conductive linear grating in a diffuse light field with stochastic parameters formed without suppressing the fluctuations.

A comparison of the images in Figs. 2 and 3 gives a basis for assuming that the anisotropy that was found is not a consequence of the regular organization of the initial data.

We observe the predicted effect only for fairly definite combinations of the radiation wavelength, the grating step, and the profile depth of its sinusoidal microrelief. The given relationships form rather wide "windows of possibilities." To illustrate what has been said, Fig. 4 shows images of the radiance diagrams of the S-polarized component of various spectral components of the blackbody radiation, scattered at the same linear reflective grating with ideal conductivity. The ratio of the total depth of the sinusoidal profile of the grating microrelief to its step is $h/d = 0.62 = Const$. Parameter "$R$" denotes the ratio between the grating step and the wavelength of the scattered component of the radiation, $R = d/\lambda$.

It can be concluded from the information presented in Fig. 4 that all the diffraction orders except the zeroth evidently participate in forming the predicted anisotropy effect. When such orders become too many, their total contribution to the cumulative scattering pattern is mutually compensated. It can thus be assumed that there is a causal connection between the appearance of this effect and the appearance of discontinuities of the phase trajectories of the photons in the regions of Rayleigh–Wood diffraction threshold anomalies [11, 12].

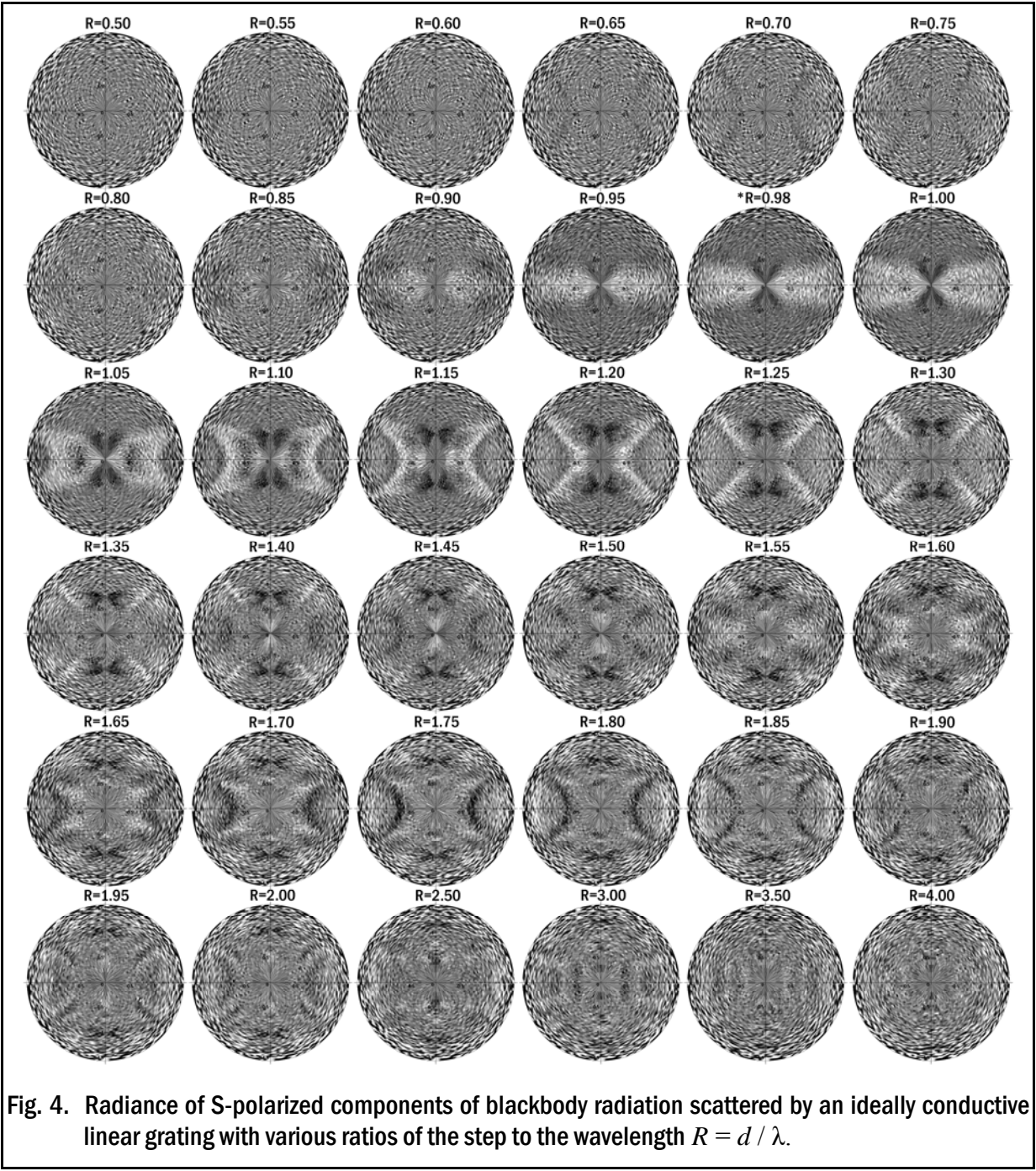

Fig. 4.  Radiance of S-polarized components of blackbody radiation scattered by an ideally conductive linear grating with various ratios of the step to the wavelength $R = d / \lambda$.

Figure 5 shows graphical images of the radiance of the polarized components of the scattered light field obtained from diffraction of blackbody radiation with temperature 5221°K at an ideally conductive linear grating (d = 544 nm, h = 337 nm).

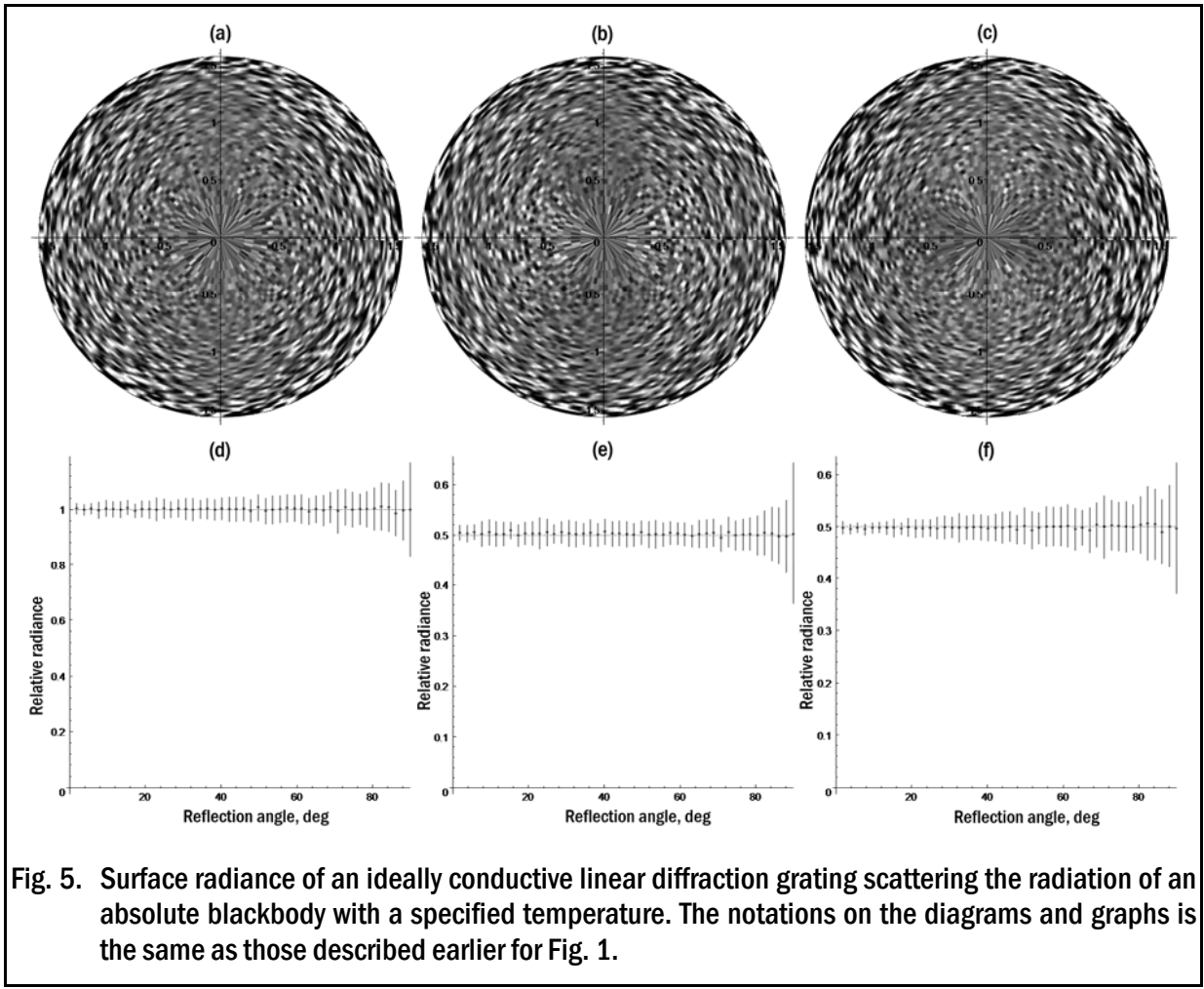

**Fig. 5.** Surface radiance of an ideally conductive linear diffraction grating scattering the radiation of an absolute blackbody with a specified temperature. The notations on the diagrams and graphs is the same as those described earlier for Fig. 1.

It can be seen from the contents of the diagrams in Figs. 5(a)– 5(c) and the corresponding graphs in Figs. 5(d)– 5(f) how the radiance gradients that characterize the separate scattered monochromatic components of the blackbody radiation (see Fig. 4) compensate each other at the total-radiance level as a consequence of the integral convolution that was carried out. The results of numerical experiments show that this compensation is observed for any temperature of the blackbody radiation and any parameters of the diffraction-grating profile and not only for those used in the example given here. It is not impossible that the isotropy of the radiance of the diffracting radiation may be strictly obligatory from the viewpoint of the second law of thermodynamics.

The described property of the convolution of the radiance imposes limitations on the methodology of using such instrumental resources as, for example, bolometric thermal-viewers in subsequent physical experiments.

## Diffraction at gratings with finite conductivity

The predicted effects can become objective if their existence is confirmed in the course of full-scale physical measurements. The interaction of a diffuse field of electromagnetic radiation with a discrete structure (a photon gas) with some ideally conductive reflecting surface was considered in the simulation experiments described earlier. Let us estimate to what extent and under what conditions the expected effects will be obtained when such a metal as aluminum, for instance, is used as the actual reflecting material.

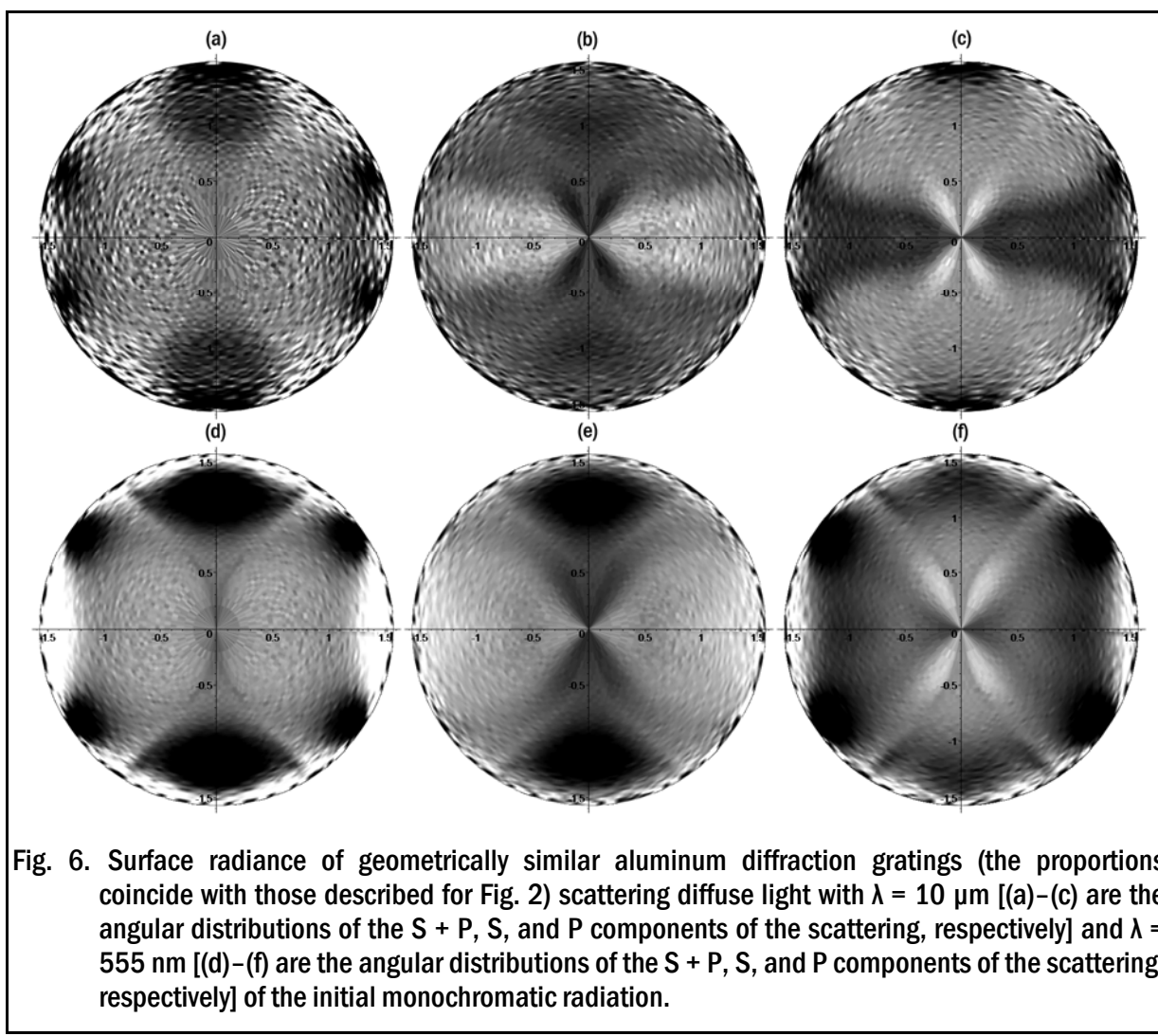

**Fig. 6.** Surface radiance of geometrically similar aluminum diffraction gratings (the proportions coincide with those described for Fig. 2) scattering diffuse light with λ = 10 μm [(a)–(c) are the angular distributions of the S + P, S, and P components of the scattering, respectively] and λ = 555 nm [(d)–(f) are the angular distributions of the S + P, S, and P components of the scattering, respectively] of the initial monochromatic radiation.

Figure 6 shows diagrams of the angular radiance distribution of the polarization components of the reflected diffuse radiation obtained for two linear geometrically similar aluminum gratings with identical relative proportions between the parameters of their microrelief profile and the wavelengths of the scattered radiation. The angular distributions in Figs. 6(a)– 6(c) display the S + P, S, and P components of the scattering of initial monochromatic radiation with wavelength 10 μm, which corresponds to the Planck maximum power of blackbody radiation at $T$ = 290°K. Figures 6(d)– 6(f) show the analogous information with respect to contents for initial monochromatic radiation having wavelength 555 nm, corresponding to the greatest physiological sensitivity of the human eye in the visible region of the spectrum.

If one now compares Fig. 2(a)– 2(c) for ideal conductivity with the information shown in Fig. 6, one can form the following conclusions:

- In the far-IR region ($\lambda = 10\ \mu m$), the character of the scattering of radiation at a grating made from a strongly conductive actual material weakly differs from the case with ideal conductivity. However, when physical experiments are being set up, the complications mentioned above in using bolometric apparatus may occur.

- In the visually observable region of the spectrum ($\lambda = 555$ nm), the absorption of radiation by the material of the diffraction grating becomes so intense that the pattern predicted for ideal conductivity in the angular distribution of the reflected light cannot be reliably recognized. Here, however, it is likely that the radiance falloff of the reflecting surface, which is especially significant in the zones of the Rayleigh–Wood resonance anomalies [13, 14], will be compensated in the state of thermodynamic equilibrium by a flux of intrinsic thermal radiation (Kirchhoff's law) in the corresponding angular directions [15]. If this kind of compensation actually exists, a positive observation of the predicted effect becomes possible in practice in the visible region.

## Conclusion

A numerically substantiated assumption has been obtained that the monochromatic components of a diffuse photon gas (for example, blackbody radiation) can acquire anisotropic polarization when it undergoes isoenergetic diffraction scattering, carried out under definite conditions. The radiance of the scattered light field in this case remains homogeneous in all directions.

In other words, blackbody radiation can acquire an inhomogeneous polarization structure when it interacts with a diffraction grating with which this radiation is in the state of thermodynamic equilibrium. The given structure, because of its greater ordering, contains information that makes it possible to visualize the topology of the diffraction grating. Meanwhile, the fact that the radiance remains isotropic eliminates the possibility of breaking the second law of thermodynamics.

The objective confirmation of this assumption gives a basis for revising the existing concept of the most probable macroscopic state of a closed system, founded on the basic postulate of statistical physics.

## Acknowledgment

The author is very grateful to Igor Golubenko, who actively participated in creating the software used in this work.

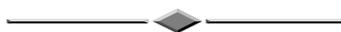